\documentclass[a4paper,11pt]{article}
\pdfoutput=1 
\usepackage{jcappub} 
\usepackage{graphicx}
\usepackage{color}
\usepackage{relsize, float, multirow}
\usepackage{array, epstopdf, hyperref, }
\usepackage{caption, subcaption, mathtools}
\usepackage{aas_macros}

\usepackage{hyperref}
\newcommand{\beq}{\begin{equation}}
\newcommand{\eeq}{\end{equation}}
\newcommand{\barr}{\begin{eqnarray}}
\newcommand{\earr}{\end{eqnarray}}

\newcommand{\xe}{x_{\rm e}}

\newcommand{\id}{{\rm d}}

\newcommand{\lsim}{\mathrel{\hbox{\rlap{\lower.55ex\hbox{$\sim$}} \kern-.3em \raise.4ex \hbox{$<$}}}}
\newcommand{\gsim}{\mathrel{\hbox{\rlap{\lower.55ex\hbox{$\sim$}} \kern-.3em \raise.4ex \hbox{$>$}}}}

\title{Cosmology with Recombination Spectrum}
\author[a]{Debajyoti Sarkar,}
\author[b]{Rishi Khatri}

\affiliation[a]{Inter-University Centre for Astronomy and Astrophysics,
  Ganeshkhind, Pune 411007, India}
\affiliation[b]{Tata Institute of Fundamental Research, Homi Bhabha Road,
  Mumbai 400005, India}

\abstract{
Precision measurement of the cosmological
recombination spectrum can provide an entire new window to look at the
early universe. We aim to quantify the information hidden in the
cosmological recombination spectrum and for this purpose we have developed
a new code following the algorithm proposed in \citep{AH10,E_c}. Our code
is  
closely based on  the COSMOSPEC \citep{cosmospec} code. We find, using
Fisher information matrix  and assuming that the foregrounds can be subtracted by using
higher or lower frequency channels and spatial information,  that going
beyond the detection will need an experiment with
sensitivity $25\times$ better compared to the proposed experiment
PIXIE. Such an experiment will be able 
to measure the cosmological parameters with a precision that is competitive with
the CMB anisotropy experiments. The best constrainted parameter is baryon energy
density, $\Omega_{\rm b}$, which can be nailed down with incredible
precision in principle. We also show that the shape of the
 hydrogen lines
 is  connected to the speed of the hydrogen recombination, with the peaks of the
 recombination lines coinciding with the peak of the recombination
 rate. In general, the shape of the lines encodes information about the
 rate of recombination as a function of redshift.}

\begin{document}

\maketitle

\section{Introduction}
Observations of the temperature and polarization anisotropies of the cosmic
microwave background (CMB) have made it possible
to determine the physics of the early universe to a very high level of
precision. The Planck CMB space mission \citep{pl13, pl15, pl18} has already extracted almost
all the critical information from the temperature anisotropies, and
 future missions like LiteBIRD\citep{litebird} , CoRE\citep{core}, PICO\citep{PICO} and  CMB-Bharat  \footnote{CMB Bharat (\href{http://cmb-bharat.in}{http://cmb-bharat.in})\label{foot:CMBBharat} } have been proposed to obtain all the cosmological information
 in the CMB polarization anisotropies. However, there is  a
 significant amount of information hidden in the  spectral distortions of
 the nearly perfect blackbody spectrum of the CMB
 \citep{1,3,6,dd1982}. Spectral distortions can be created due  to energy
 injection in the early universe, at redshifts $z\lesssim 2\times 10^6$, resulting in $\mu-$type, $y-$type,
 $i$-type, 
 or non-thermal relativistic  distortions \citep{Chluba:2011hw,ks2013,c2013,AK2018,sandeep}. 
There are many unavoidable  sources of  spectral distortions in standard
cosmology \citep{Chluba:2011hw, sunyaev_khatri_review_2013} such as  silk
damping \cite{3,daly1991,hss94,cks2012,ksc2012b},  y-distortion from the reionization era, and
the cosmological recombination spectrum \citep{zks68,peebles68,d1975,sss2000,dg2005,cs2006,rcs2006,cs2006b,ki2006,Wong:2005yr,ws2007,crs2007,cs2007,h2008,ki2008,sh2008,sh2008b,sh2008c,gd2008,kiv2008,rcs2008,cs2008,cs2009b,hf2009,cs2009c,cs2010,cs2010b,c2010,gh2010,agh2010,cfs2012}.

We will focus on the cosmological recombination spectrum in this paper. The primordial hydrogen and helium recombinations \citep{zks68,peebles68} create few 
photons per atom \citep{cs2006b}, which lead to a small distortion of
$\Delta I_{\nu}/I_{\nu} \approx 
10^{-9}$ in the CMB, where $I_{\nu}$ is the blackbody CMB intensity and
$\Delta I_{\nu}$ is the intensity of the recombination lines.  The cosmological recombination spectrum
contains a wealth of information and is in particular not constrained by the
cosmic variance which limits the precision with which we can measure the
cosmological parameters with CMB anisotropies. In particular,
it  may provide an excellent measurement of the primordial helium
abundance, uncontaminated by stellar contributions. The
recombination spectrum can also be a probe of non-standard energy
injections such as dark-matter annihilations \citep{c2010}.

In 1991, the observations of the cosmic microwave background (CMB) spectrum
with Cosmic Background Explorer-Far Infrared Absolute Spectrophotometer
(COBE/FIRAS) \citep{cobe} showed that the CMB spectrum is close to a
perfect black-body, with fractional deviations less than few $\times
10^{-5}$.  Unlike the CMB anisotropies, which have seen precision improve
by many orders of magnitude,  the measurement of the CMB spectrum has not
been updated since then. However, over the last 25 years the technology has improved
considerably \cite{fm2002}. In recent years, various missions have been proposed
\citep{pixie, prism} to measure CMB spectral distortions. These proposed
instruments are expected to detect the spectral deviations of the order of $\Delta I_\nu/
I_\nu \sim 10^{-8}$ to $10^{-9}$, i.e. 3 to 4-orders of
magnitude improvement over COBE-FIRAS. With this kind of sensitivity, it is
possible to detect the recombination lines if we can separate them
from foreground emissions. Since the recombination spectrum is expected to
be isotropic and it has a very characteristic shape, this separation from the
foregrounds should be possible in principle\citep{Rao}, although a clear demonstration
of how accurately  we can measure the cosmological recombination spectrum
is pending and will require new foreground separation algorithms to be developed.

In this paper, we will not tackle the difficult question of foreground
separation but rather attempt to quantify the information content of the
cosmological recombination spectrum. We will assume that the frequency
range $30~{\rm GHz}\le \nu \le 600~{\rm GHz}$ can be cleaned of foregrounds
using frequency channels at $\nu \le 30~{\rm GHz}$ and $\nu \ge 600~{\rm
  GHz}$ as assumed in the PIXIE proposal \cite{pixie}. We want to ask the
question what is the minimum improvement of sensitivity that is needed over
PIXIE type experiment in order for the cosmological recombination spectrum
to be competitive with the CMB anisotropies for measurement of the
cosmological parameters. We also want to ask, given net total sensitivity
of the experiment, whether it is more advantageous to divide the frequency
channels into finer resolution channels with smaller sensitivity in each
channel, or whether we want to have broader channels with higher
sensitivity in each channel. This question has practical implications for the design
of future experiments, since increasing the frequency resolution in a PIXIE
type experiment increases the physical size of the experiment
\cite{pixie}. We also want to explore the  complementarity of  the
recombination spectrum with the CMB anisotropy spectrum.

Computing the recombination spectrum requires  computation of the
populations of the excited states at each redshift, finding out
corresponding the photon emission or absorption and finally solving the radiative transfer equation. High-precision 
computation of spectra requires accounting for excited states up to principal quantum
number $n_{\max}$ of a few hundred and resolving the angular momentum sub-states. With the
standard multilevel atom method, this will require a large computational
time \citep{Seager:1999km,cs2006b, cs2006, ki2006,cs2007,crs2007,cs2010}.
We follow the effective conductance method developed by \citep{E_c}, which
bypasses this computational problem by dividing the problem into two parts:
the cosmology independent but computationally expensive part which only depends
on atomic physics and temperature and the cosmology dependent but
computationally inexpensive part. The cosmology independent
part depends only on temperature and hence needs to  be computed once and
for all and tabulated
 as a function of  matter and radiation temperature, to be used again and again
 as needed \citep{E_c}. The cosmology dependent part needs to be calculated
 every time we change the cosmological parameters.  However, computing this
 part is fast, and hence, the total computation time is much smaller
 resulting in a fast and precise multi-level code  without any fudge factors.

 We closely follow the implementation of  COSMOSPEC and refer the reader to
 \citep{cosmospec} for details. We briefly review the essential aspects of
 the calculation below.

\section{Computation of the cosmological recombination spectrum} \label{sec:equations}

In the optically thin limit, and in the absence of any source, the
occupation number or the phase-space density of photons, $I_{\nu}/\nu^3$
can be assumed to be conserved, $\id (I_{\nu}/\nu^3)\id t=0$, where
$I_{\nu}$ is the intensity, $t$ is
the proper time, and the derivative is a total derivative. The only
dependence on  time of the intensity in this case is  implicit and comes solely due
to the redshifting of the photons due to the expansion of the Universe. The solution
to the above equation in the expanding Universe gives a simple relation for the
intensity observed at redshift $z'$, relating it to intensity at an earlier
redshift $z$,
\beq
I_{\nu}(z) = \left(\frac{\nu}{\nu'}\right)^3
I_{\nu'}(z'),\label{eq:free-redshifting}.
\eeq
where $\nu'=\nu(1+z')/(1+z)$ is the redshifted frequency at $z'$.
 During the recombination era, ions and electrons recombine to
 form atoms, emitting  bound-bound and free-bound transition photons in the process which
 form the recombination spectrum. These photons form the source term or the
 emissivity $j_{\nu}$ in the radiative transfer equation,
\barr
\frac{d}{dt}\left(\frac{I_{\nu}}{\nu^3}\right) = c\frac{j_{\nu}}{\nu^3}.
\earr
Our main goal is to calculate this emissivity during the recombination epoch.

\subsection{The effective conductance method} \label{sec:conductance}
 In the original three level atom of \cite{zks68,peebles68}, only the first
 excited level of hydrogen ($n=2$) was resolved, and all higher levels were assumed
 to be in thermal equilibrium with the $n=2$ level. In the three-level
 atom, therefore, the only photon emission comes from de-excitation from
 $n=2$ to $n=1$ level.  The $Ly-\alpha$ photons, produced in $2p-1s$
 transition  mostly get captured by
the recombining neutral hydrogen atoms, due to the high optical depth of the line, and only a very small fraction escape.
  This makes the very weak  $2s-1s$ double photon decay also
 very important. The $2s-1s$ channel contributes significantly to the
 recombination process and hence to the  recombination
 spectrum \citep{cs2006, cs2008, h2008, sh2008b} and needs to be taken into
 account. In general, we need to take into account all allowed transitions
 between all excited levels as well as weak transitions, such as 2-photon decays, from excited
 states to the ground state in a multi-level atom. The effective multi-level approach \cite{AH10}
 divides the excited levels of atoms into interface states, $i$, defined as the states from which direct transitions to the ground
 state are taken into account. The populations of these levels, $x_i$, are
 explicitly followed. All other states are interior states  from which direct transitions to the ground
 state are not important or not allowed. The effective multilevel atom
 takes the effect of these interior states also into account almost
 exactly, without explicitly solving for these states.

The effective conductance method  extends
 this approach to compute the cosmological recombination spectrum in an
 efficient manner \citep{E_c}. This is accomplished by realizing that in
 Saha equilibrium, the net emission would be balanced by
 absorption. Therefore, the net emissivity, $j_{\nu}$,  will depend only on
 the 
 departure of the level populations from the Saha equilibrium. 
The effective multilevel atom of \cite{AH10} only keeps track of the
populations in the 
interface levels.  It was realized in \cite{E_c} that the departures from
the Saha equilibrium of all levels $\Delta x_{n\ell}$, where $n,\ell$ are
the quantum numbers of the level, including interior
states  can be solved in
terms of the level populations of interface levels, without explicitly
following all interior states. Thus the  net transition rates
between two levels are proportional to departure of interface states from
equilibrium, $\Delta x_{i}$, with the
constant of proportionality, the effective conductance, only a function of matter and radiation temperatures and atomic
physics and does not depend on cosmology or the actual level
populations. These effective conductances 
can therefore also be computed and tabulated once and for all. We refer the
reader to \cite{E_c} for more details.

\subsection{He recombination and feedback} 
 It is essential to model helium recombination  to compute the spectral
 distortion due to total recombination precisely. Since the ionization
 energy of helium is greater than hydrogen, helium recombination happens at
 much higher redshift compared to hydrogen recombination. The photons from
 the He recombination were more energetic, and they had to travel a much
 higher distance compared to photons originated due to hydrogen
 recombination. The helium photons interact with hydrogen as well as He
 atoms.  Since the number density of helium atoms is $\approx 8\%$ of that
 of hydrogen,  the intensity of the helium recombination spectrum  is also
 $\approx 8\%$ of that of hydrogen. Photons emitted during the
 recombination phases of helium can ionize and excite the hydrogen at a
 later time. He II recombination photons can affect  HeI and hydrogen
 recombination and HeI recombination photons can affect hydrogen
 recombination. These ionizations and excitations change the recombination
 spectra. We followed \citep{cosmospec} to compute and take into account
 feedback effects. We take  He levels into account up to principle quantum
 number of 100.

\section{Cosmology with the recombination spectrum}
\subsection{Comparison with the COSMOREC}
\begin{figure}[!btp]
\centering
  \begin{subfigure}[b]{0.3\textwidth}
\includegraphics[scale=0.3]{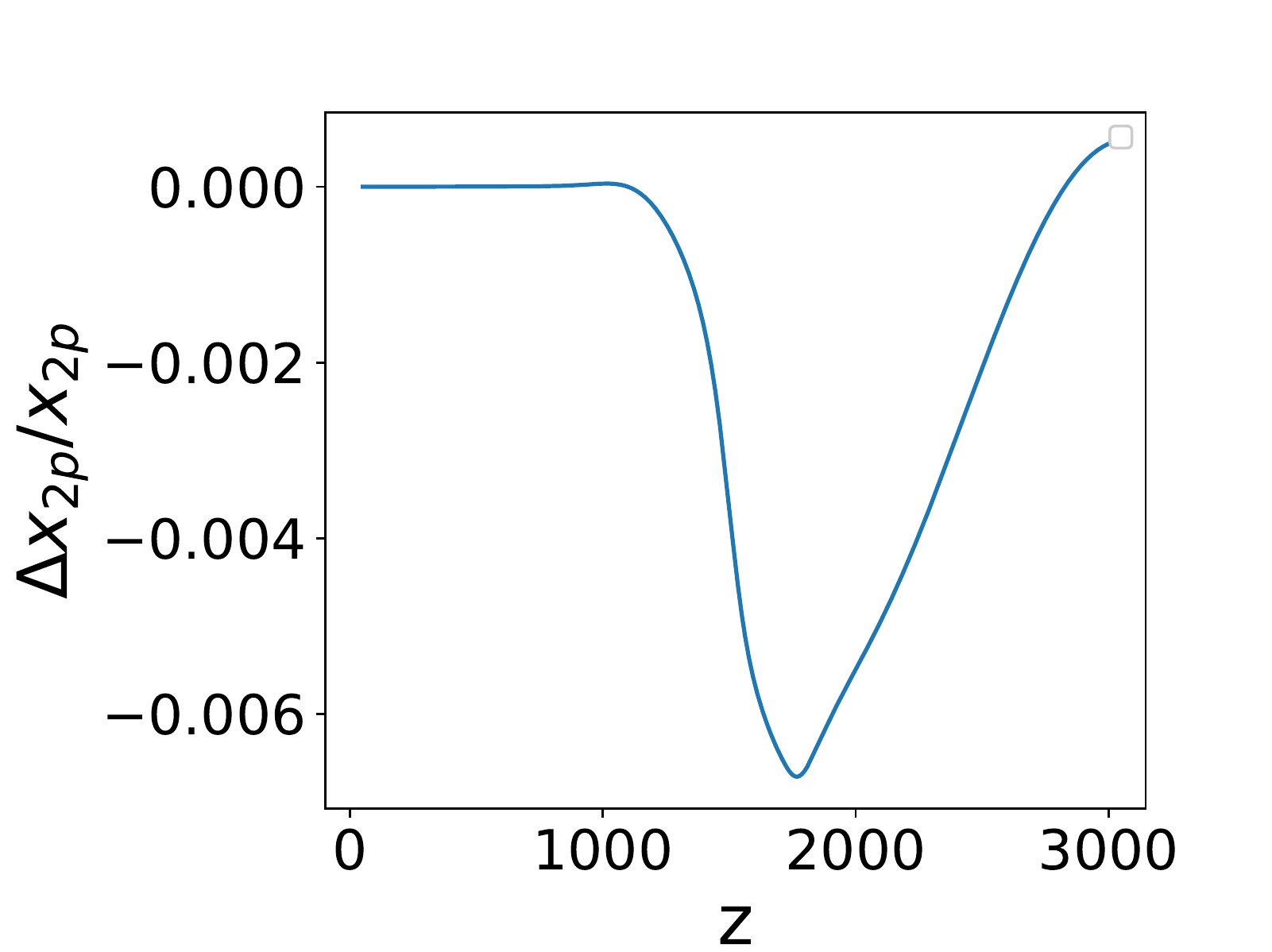}
    \caption{}
  \end{subfigure}
  \begin{subfigure}[b]{0.3\textwidth}
\includegraphics[scale=0.3]{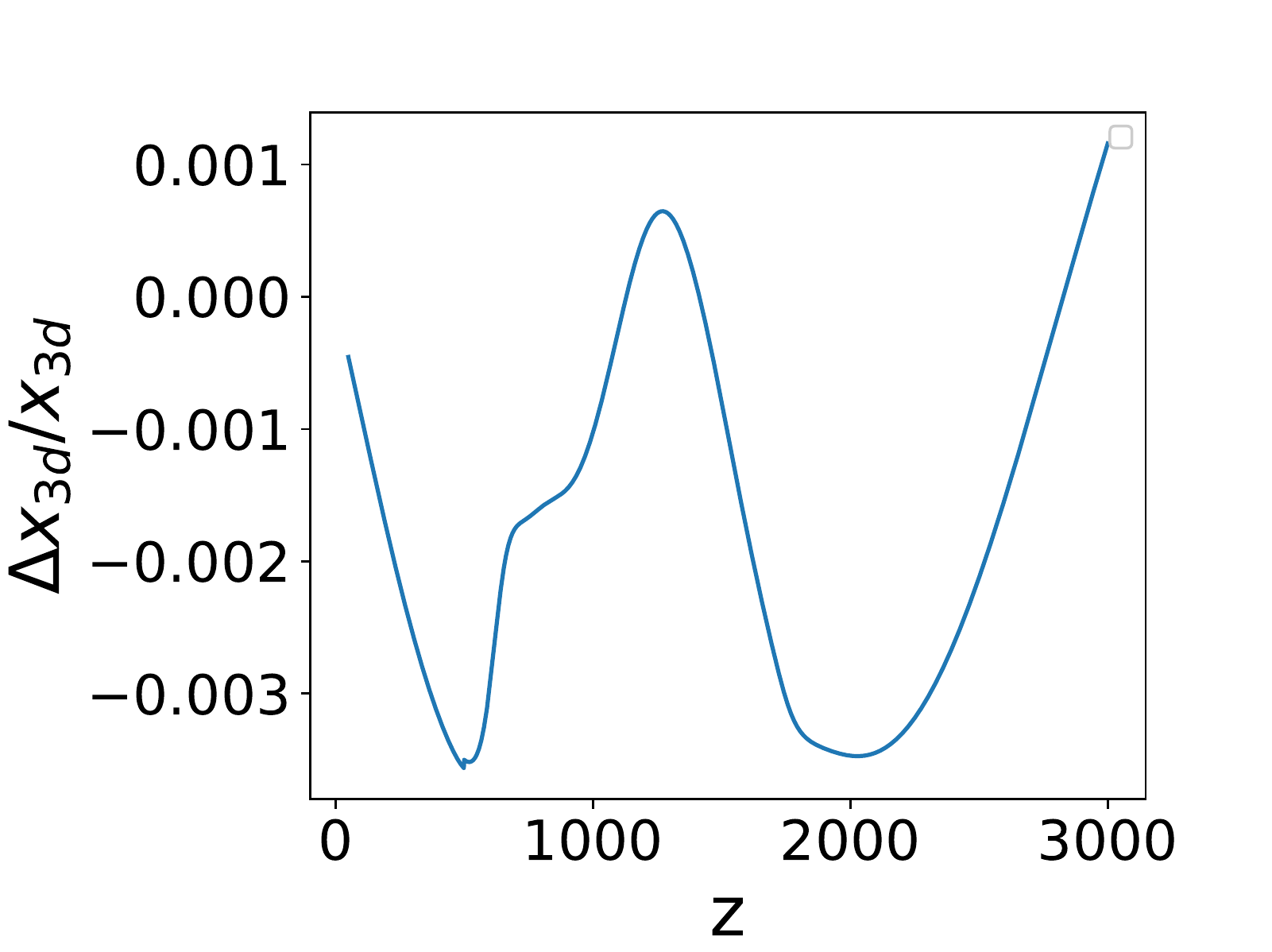}
    \caption{}
  \end{subfigure}
  \begin{subfigure}[b]{0.3\textwidth}
\includegraphics[scale=0.3]{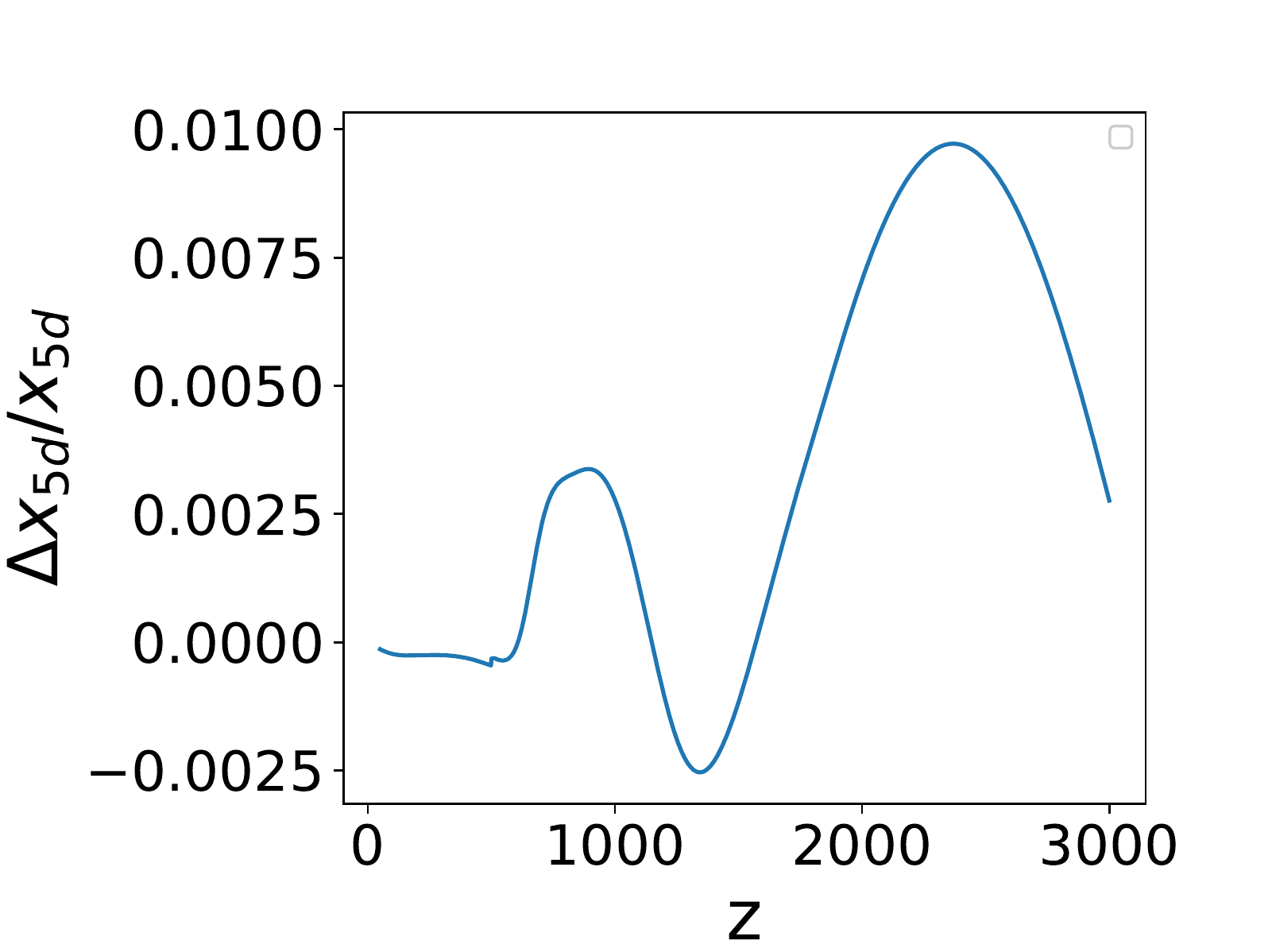}
    \caption{}
  \end{subfigure}
\caption{(a) The fractional difference between 
   our code and COSMOREC for level populations, $\Delta x_i/x_i$: (a) 2p
   (b) 3d (c) 5d}\label{fig:comp}
\end{figure}

\begin{figure}
\centering
\resizebox{\hsize}{!}{\includegraphics{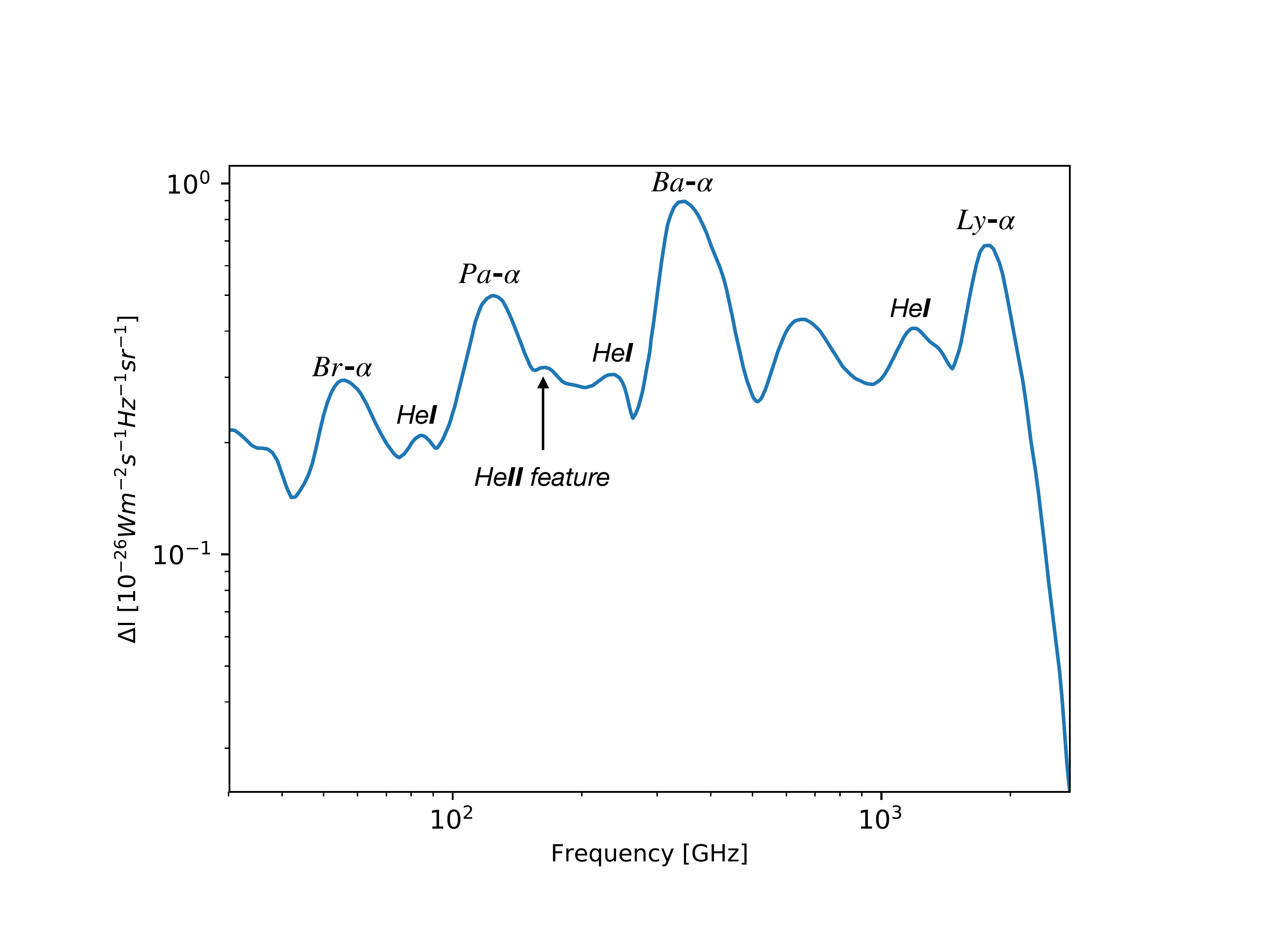}}
\caption{ The cosmological recombination spectrum}\label{fig:spec}
\end{figure}

To check the accuracy of our implementation of the recombination
calculation we compare with the publicly available COSMOREC code.  The
COSMOREC code \citep{cosmorec} computes the recombination history and has
the option to resolve first few levels of hydrogen and helium. We have done
detailed comparison of first 10 levels of hydrogen and first 5 levels of
helium and found agreement at the level of few $\%$. We show the comparison of the 
  population levels of excited hydrogen states as computed by
COSMOREC\citep{cosmorec} and our code in Fig. \ref{fig:comp}. for 3 excited hydrogen
levels. The full recombination spectrum, including contribution from helium,
is shown in Fig. \ref{fig:spec}.

\subsection{Fisher Matrix analysis}
\begin{table}[hbtp]

\begin{center}
\caption{Experiment sensitivities for Fisher forecasts. For all experimental
configurations we use the minimum and maximum frequency of $30~$GHz and
$600~$GHz respectively. In addition for the same per channel sensitivity we
consider 15GHz channels (same as PIXIE) and also 5GHz channels, effectively
increasing the total sensitivity by $\sqrt{3}$ in the latter case over the
former. We thus have total  $39$ channels for 15GHz channel widths  and $115$
  channels for 5GHz channel widths. PIXIE sensitivity is shown just for reference.}
\label{tbl:exp}
\begin{tabular}{ |c|c|c|c|c|c|c| } 
 \hline
 Proposed Experiment & PIXIE & $25\times$ PIXIE&$\sqrt{3}\times 25\times$PIXIE& $100\times$ PIXIE\\
 \hline 
Sensitivity per channel ($Jy/sr$) &  5 & 0.2  & 0.115 & 0.05  \\ 
 \hline
\end{tabular}
\end{center}
\end{table}
The Fisher information matrix for a series of independent data points/observables
$f_b$, which are function of parameters $p_i$ with Gaussian uncertainties
$\sigma_b$ for each data point is defined as
\begin{equation}
F_{jk}=\sum _b\frac{1}{\sigma_b ^2} \frac{\partial f_b}{\partial p_j} \frac{\partial f_b}{\partial p_k}
\end{equation}
The covariance matrix is the inverse of the Fisher matrix. The Fisher
matrix provides a best-case scenario for the ability of a a set of  experiments to constrain
parameters.  For our case, the observables
are the recombination spectrum intensities in different frequency
bands. The recombination spectrum is a function of  the cosmological
parameters the baryon density parameter ($\Omega_b$),  the cold dark matter
density parameter ($\Omega_c$), the mass fraction of helium ($ Y_p$),  the
Hubble constant ($H_0$, $h\equiv H_0/100$), and the effect relativistic  degrees of freedom 
($N_{\rm eff}$). The Fisher matrix depends on the values of fiducial parameters where
we evaluate the derivative. We use Planck 2018 \cite{pl18} TT+lowE best fit parameters as our
fiducial cosmology. The intensity  in each frequency channel is  an
independent data point  $b$.  We also want
to compare the forecasts with CMB anisotropy as well as give the combined
forecasts for recombination spectrum and the CMB anisotropies to explore
their complementarity. We use the
Planck results as the reference case for the anisotropies, although we
should expect improvements in anisotropy measurements, especially
polarization anisotropies, with future experiments \textsuperscript{[\ref{foot:CMBBharat}]} \citep{core, PICO, litebird} .We use the covariance matrix from the Planck Likelihood data
from Planck Legacy
Archive.\footnote{\href{https://pla.esac.esa.int/\#home}{https://pla.esac.esa.int/\#home}}
For the joint forecast, the combined Fisher matrix is then just the sum of
two Fisher matrices. 

\begin{figure}[hbtp]
\centering
\includegraphics[scale=0.4]{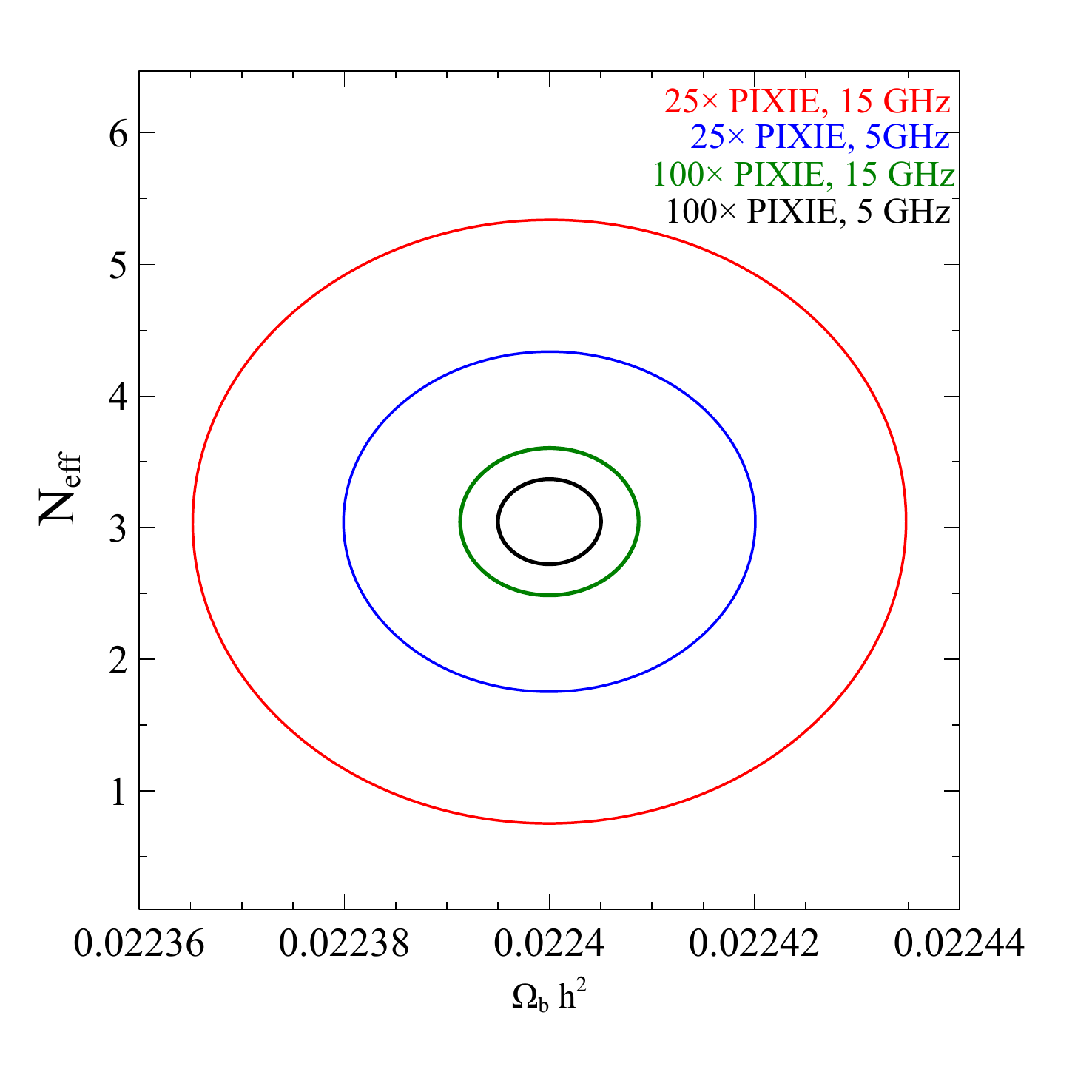}

\caption{Marginalized  68 percent confidence constraints for $\Omega_b h^2$
  and $N_{\rm eff}$ for different experiment configurations. We label the
  improvement in sensitivity per channel over PIXIE sensitivity per channel
  as $N \times$PIXIE. The minimum and maximum frequency channels are at
  $30$ and $600$ GHZ respectively. This given $39$ channels for 15GHz channel widths  and $115$
  channels for 5GHz channel widths.}
\end{figure}

To obtain the error ellipses for any subset of parameters, we need to
marginalize the Fisher matrix.   Let  the full parameter set  be
represented by $\vec{i}$ and a subset we are interested in by  $\vec{j}$
and the rest of parameters by $\vec{k}$. The Fisher matrix $F^\prime$ after marginalization over $\vec{k}$ can be expressed as,
\barr
F^\prime =F_{jj}-F_{jk}F^{-1} _{kk}F_{kj}
\earr
Inverting $F^\prime$ gives the marginalized covariance matrix for 
parameters of interest.

\begin{figure}[hbtp]
\centering
\includegraphics[scale=0.6]{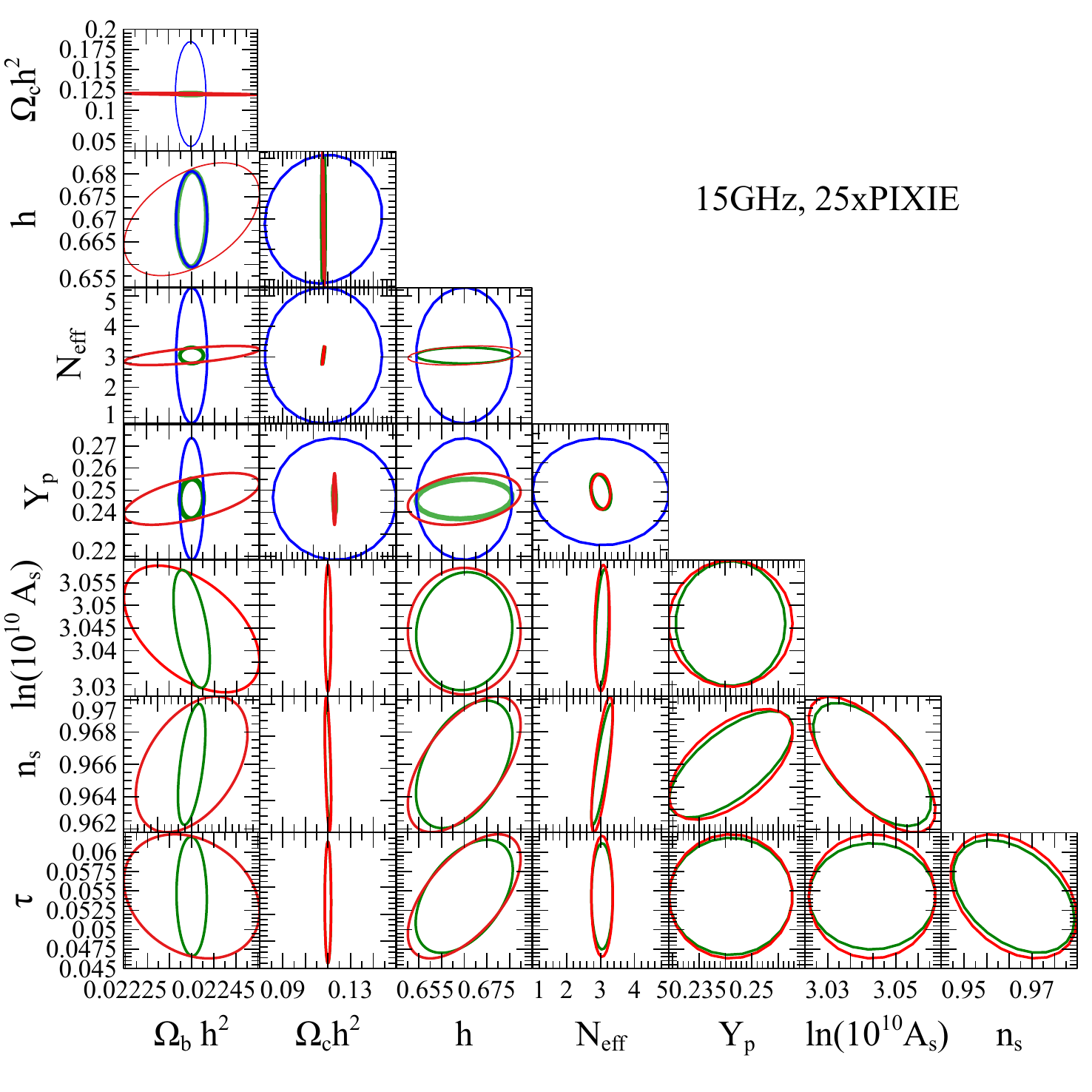}

\caption{Constraints for a PIXIE type experiment with 15GHZ wide channels
  but $25\times$ more sensitivity per channel (blue). We also show constraints
  when combined with Planck (green) and Planck 2018 constraints (red).}\label{fig:25x15}
\end{figure}

\begin{figure}[hbtp]
\centering
\includegraphics[scale=0.6]{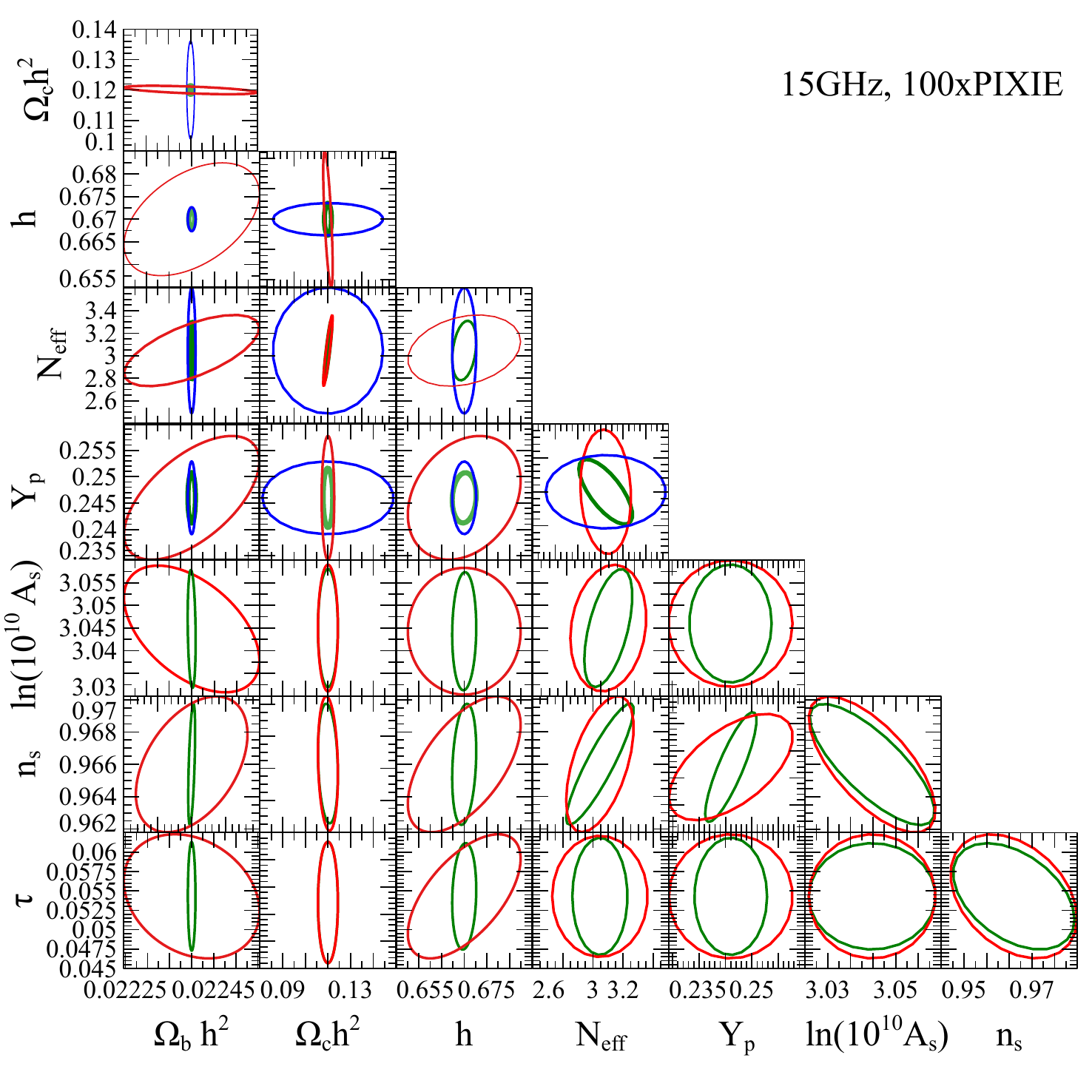}

\caption{Same as Fig. \ref{fig:25x15} but with 15GHz channels and $100\times
  $PIXIE sensitivity per channel.}\label{fig:100x15}
\end{figure}

\begin{figure}[hbtp]
\centering
\includegraphics[scale=0.6]{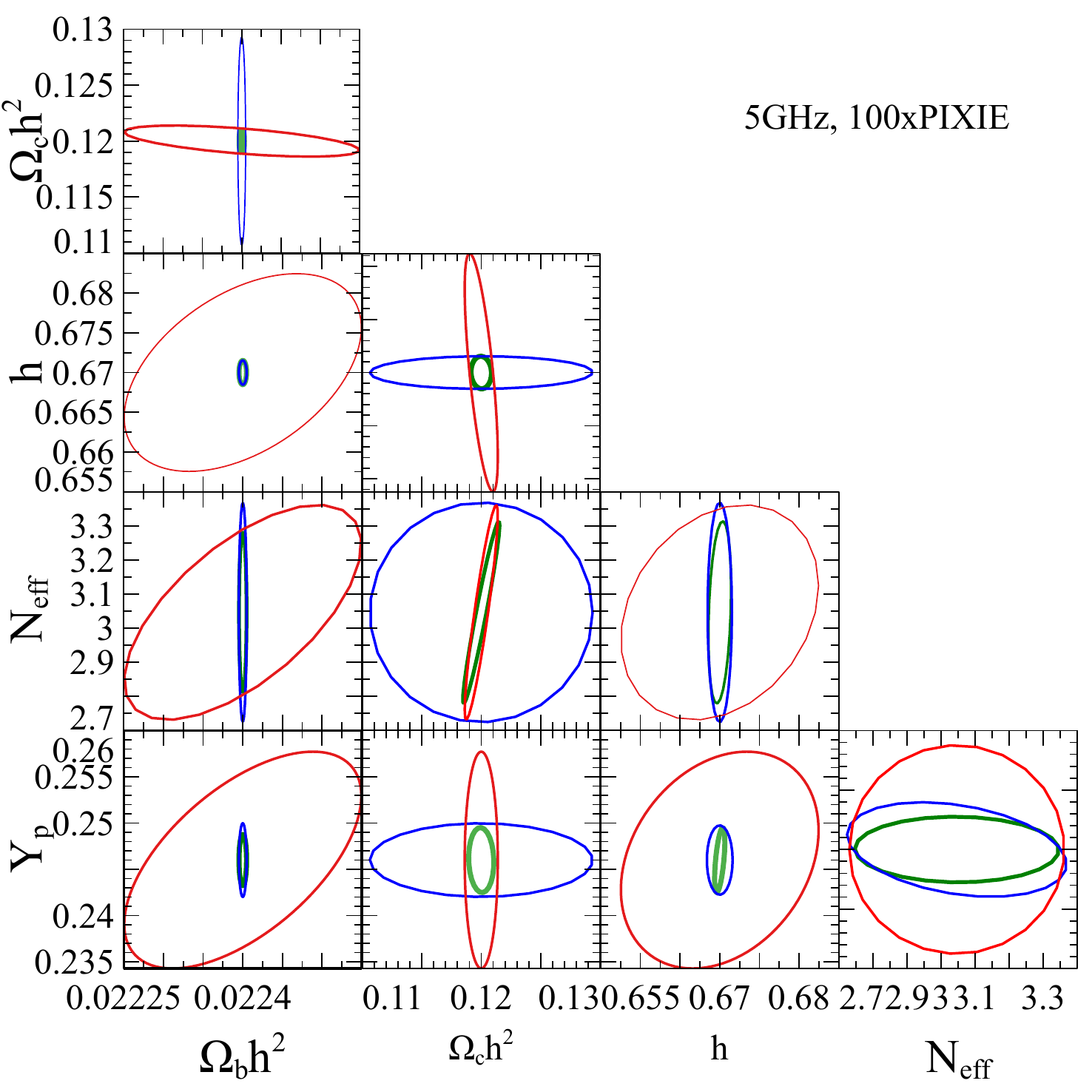}

\caption{Same as Fig. \ref{fig:25x15} but with 5GHz channels and $100\times
  $PIXIE sensitivity per channel. }\label{fig:100x5}
\end{figure}

\begin{figure}[hbtp]
\centering
\includegraphics[scale=0.6]{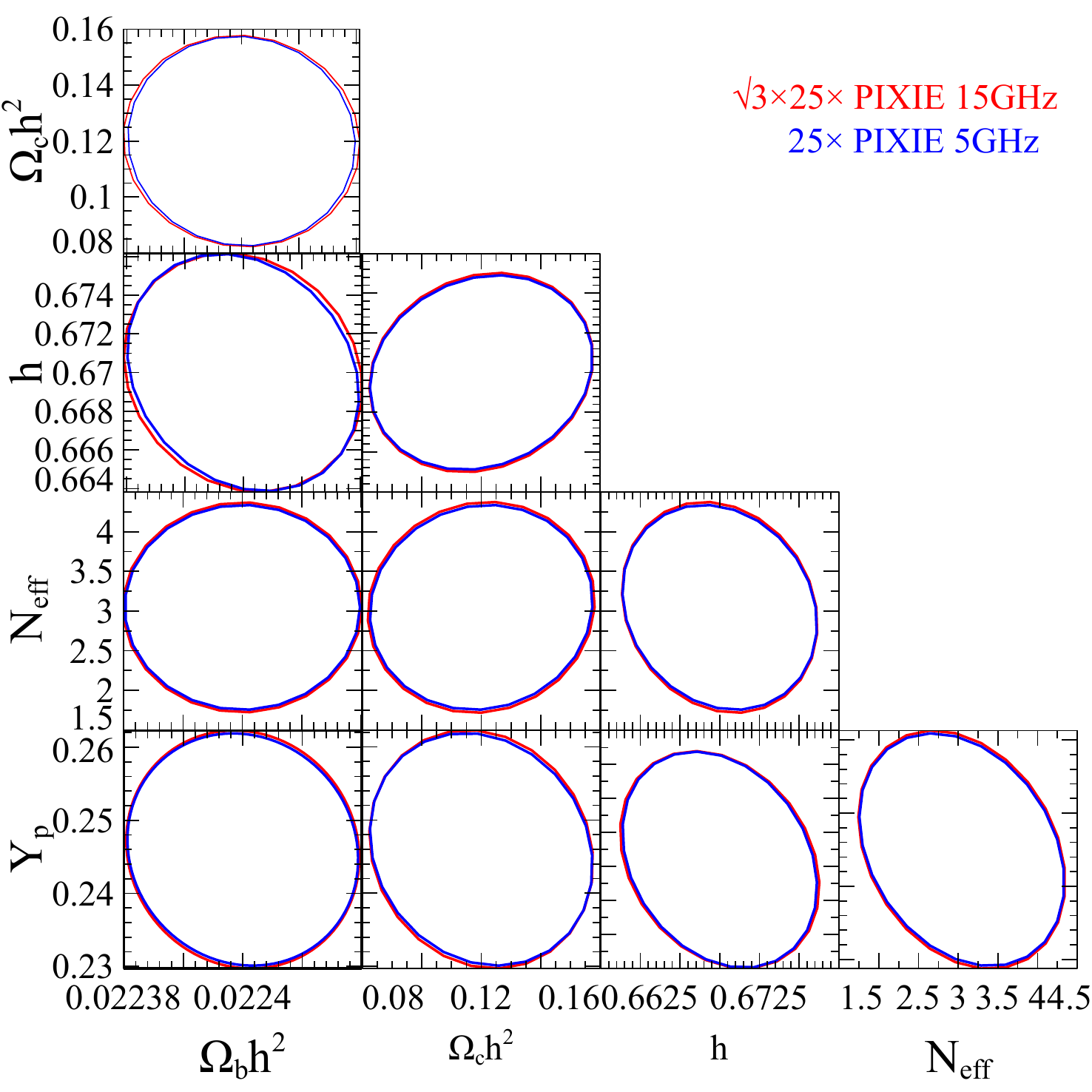}

\caption{Same as Fig. \ref{fig:25x15} but with 5GHz channels and $25\times
  $PIXIE sensitivity and 15GHz channels and $\sqrt{3}\times25\times$PIXIE
  per channel sensitivity. }\label{fig:25x53}
\end{figure}

\begin{figure}[hbtp]
\centering
\includegraphics[scale=0.6]{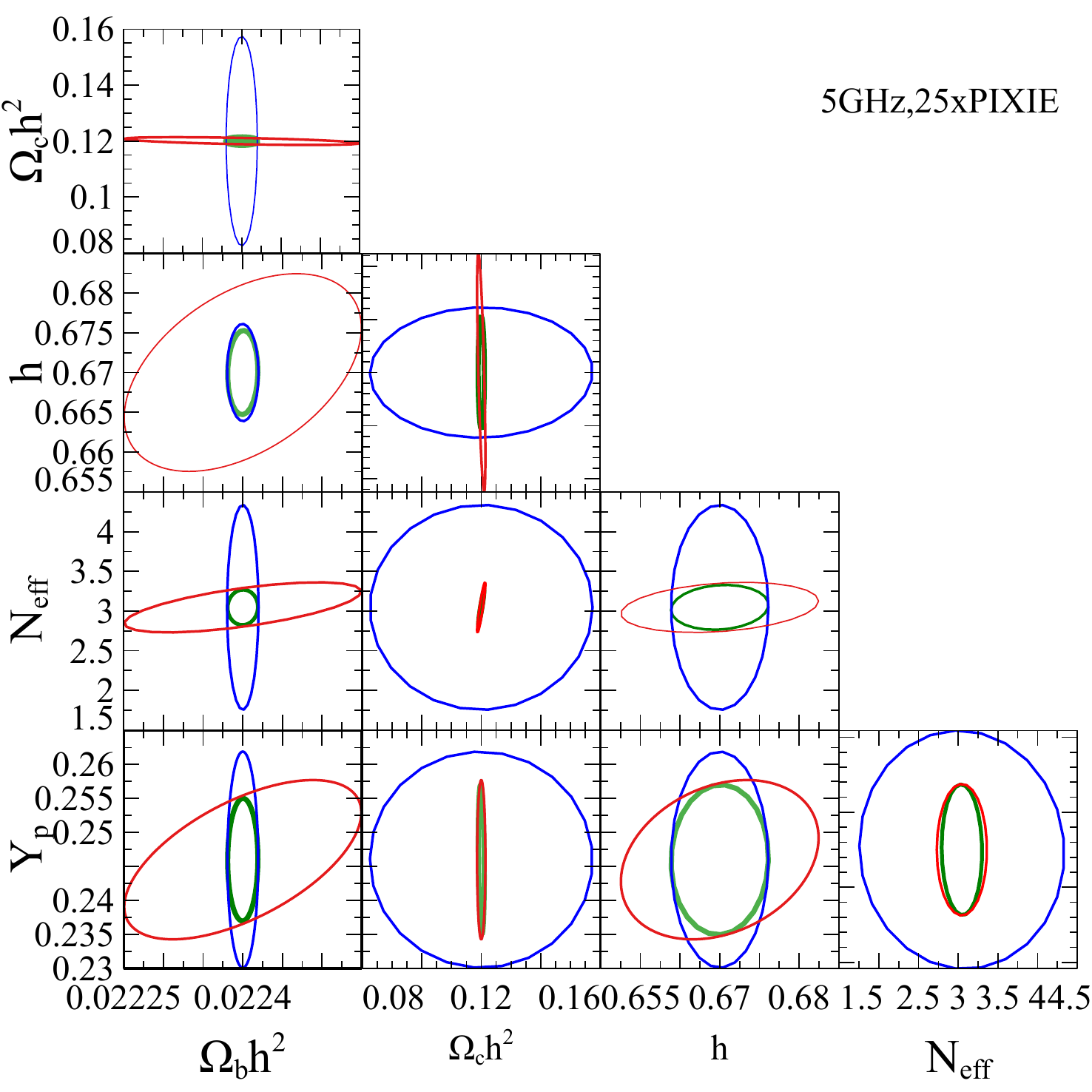}

\caption{Same as Fig. \ref{fig:25x15} but with 5GHz channels and $25\times
  $PIXIE per channel sensitivity. }\label{fig:25x5}
\end{figure}

We consider several experiment configurations, shown in Table \ref{tbl:exp}, with PIXIE \cite{pixie}
proposal as the reference.
A PRISM-like \citep{prism} instrument with $\approx 8\times$  better sensitivity
compared to PIXIE  will be able to just detect the recombination
spectrum. However we want to go beyond that and consider precision
measurement of cosmological parameters from recombination
spectrum. Therefore the minimum sensitivity of the experiments we consider
is a factor of $25$ better compared to PIXIE. In
principle, since there is no cosmic variance, we can keep on increasing the
sensitivity until we are limited by the foregrounds. On the other hand,
foregrounds can be subtracted more and more accurately by increasing the
number of frequency channels. At the other extreme, we consider an
ultra-futuristic experiment with $5~$GHz channels and a sensitivity per
channel of $0.05~{\rm Jy/sr}$, i.e. $100\sqrt{3}$ times more combined sensitive than
PIXIE. We will 
refer to per channel sensitivity everywhere with reference to PIXIE. Therefore, a 25XPIXIE experiment with
5GHZ channels has $\sqrt{3}$ better sensitivity compared to a 25XPIXIE
experiment with 15GHz channels, assuming same minimum and maximum
frequencies. The exact sensitivity values are given in Table \ref{tbl:exp}.

The Fisher forecasts are shown in  Figs. \ref{fig:25x15}-\ref{fig:25x5} as
2-parameter contour plots and in Tables \ref{tbl1} and \ref{tbl2} as
1-$\sigma$ expected errorbars on different parameters. We
see that a factor of $25$
sensitivity would nail down the baryon density, giving a factor of 5
improvement over Planck. The cosmological recombination spectrum
is less sensitive to the other cosmological parameters but in principle
still deliver competitive constraints. In particular, it is  possible to
estimate helium fraction, and Hubble parameter 
and $h_0$ better than the Planck \citep{pl13,pl15,pl18} but it will require
sensitivity that is a factor of $100$ better compared to Planck.  When combining with
Planck, we see that there is small improvement in $N_{\rm eff}$ and
Hubble parameter $h$ due to their small degeneracy with $\Omega_b
h^2$.  However, first and foremost, the cosmic recombination spectrum is a
precise \emph{baryon-meter}. There is almost no effect on the perturbative parameters, the amplitude
of primordial curvature perturbation ($A_{\rm s}$), its spectral index 
($n_{\rm s})$, and optical depth to reionization $\tau_{\rm ri}$. In
general, any parameter which has degeneracy with baryon density would
benefit from the cosmological recombination spectrum.

In Fig. \ref{fig:25x53} we compare two experiments with the same total
combined sensitivity but different frequency resolution. The two
experiments are almost indistinguishable indicating that the frequency
resolution does not affect the precision of cosmological parameters. The
frequency resolution of the experiment should therefore be determined by
foreground cleaning requirements. In general, as the total  sensitivity increases,
we will need better foreground cleaning and hence better frequency
resolution. We leave a detailed study of this very important aspect to
future work.

\begin{table}[hbtp]
\begin{center}
\caption{Constraints on standard cosmological parameters from Planck data
  and different experiment configurations defined in Table \ref{tbl:exp}.
}\label{tbl1}
\begin{tabular}{ |c|c|c|c|c|c|c| } 
 \hline
 Parameters & Planck & $25\times$ PIXIE&Planck$+25\times$PIXIE& $100\times$ PIXIE& Planck$+100\times$PIXIE \\
 \hline 
$ \Omega_bh^2$   & $0.00015$  & $0.000031$ & $ 0.000031$ & $ 0.00000707$ & $ 0.00000706$ \\ 
 $\Omega_c h^2$ & $0.0014$ & $0.065$&$0.00138$&$ 0.01625$ & $0.00135$\\ 
$ N_{eff}$   & $0.31$  & $2.202$ & $0.264$ & $0.56$ &$0.26$ \\ 
 $Y_p$ & $ 0.0117$ &$0.032$ & $0.0089$ & $0.0078 $ & $0.00547$ \\  
 $ h$   & $0.012$  & $ 0.0109$ & $ 0.01$& $ 0.00271$ & $ 0.00193$ \\ 
  $ n_{\rm s}$   & $0.0042$  &  & $ 0.00398$& $ $ & $0.00394$ \\ 
    $ln(10^{10} A_{\rm s})$   & $0.014$  &  & $ 0.0131$& $ $ & $0.01305$ \\ 
      $ \tau_{\rm ri}$   & $0.0073$  &  & $ 0.00706$& $ $ & $0.007$ \\

 \hline
\end{tabular}
\end{center}
\end{table}

\begin{table}
\begin{center}
\caption{Constraints on standard cosmological parameters from Planck data
  and different experiment configurations defined in Table \ref{tbl:exp}.
}\label{tbl2}
\begin{tabular}{ |c|c|c|c|c|c|c| } 
 \hline
 Parameters & Planck & $25\times$ PIXIE & Planck$+25\times$PIXIE& $100\times$ PIXIE& Planck$+100\times$PIXIE \\
 \hline 
$ \Omega_bh^2$   & $0.00015$  & $0.00001789$ & $ 0.0000178$ & $ 0.00000447$ & $ 0.0000044$ \\ 
 $\Omega_c h^2$ & $0.0014$ & $0.0373$&$0.0012$&$ 0.0093$ & $0.0011$\\ 
$ N_{\rm eff}$   & $0.31$  & $1.29$ & $0.29$ & $0.32$ &$0.25$ \\ 
 $Y_p$ & $ 0.0117$ &$0.0184$ & $0.008$ & $0.0045$ & $0.0039$ \\  
 $ h$   & $0.012$  & $ 0.006293$ & $ 0.006285$& $ 0.00153$ & $ 0.00152 $ \\ 
  $ n_{\rm s}$   & $0.0042$  &  & $ 0.00396$& $ $ & $0.00394$ \\ 
    $ln(10^{10} A_{\rm s})$   & $0.014$  &  & $ 0.01301$& $ $ & $0.013$ \\ 
      $ \tau_{\rm ri}$   & $0.0073$  &  & $ 0.007$& $ $ & $0.007$ \\
 \hline
\end{tabular}
\end{center}
\end{table}

\subsection{The shape and position of the lines}

\begin{figure}[hbtp]
\centering
\includegraphics[scale=0.4]{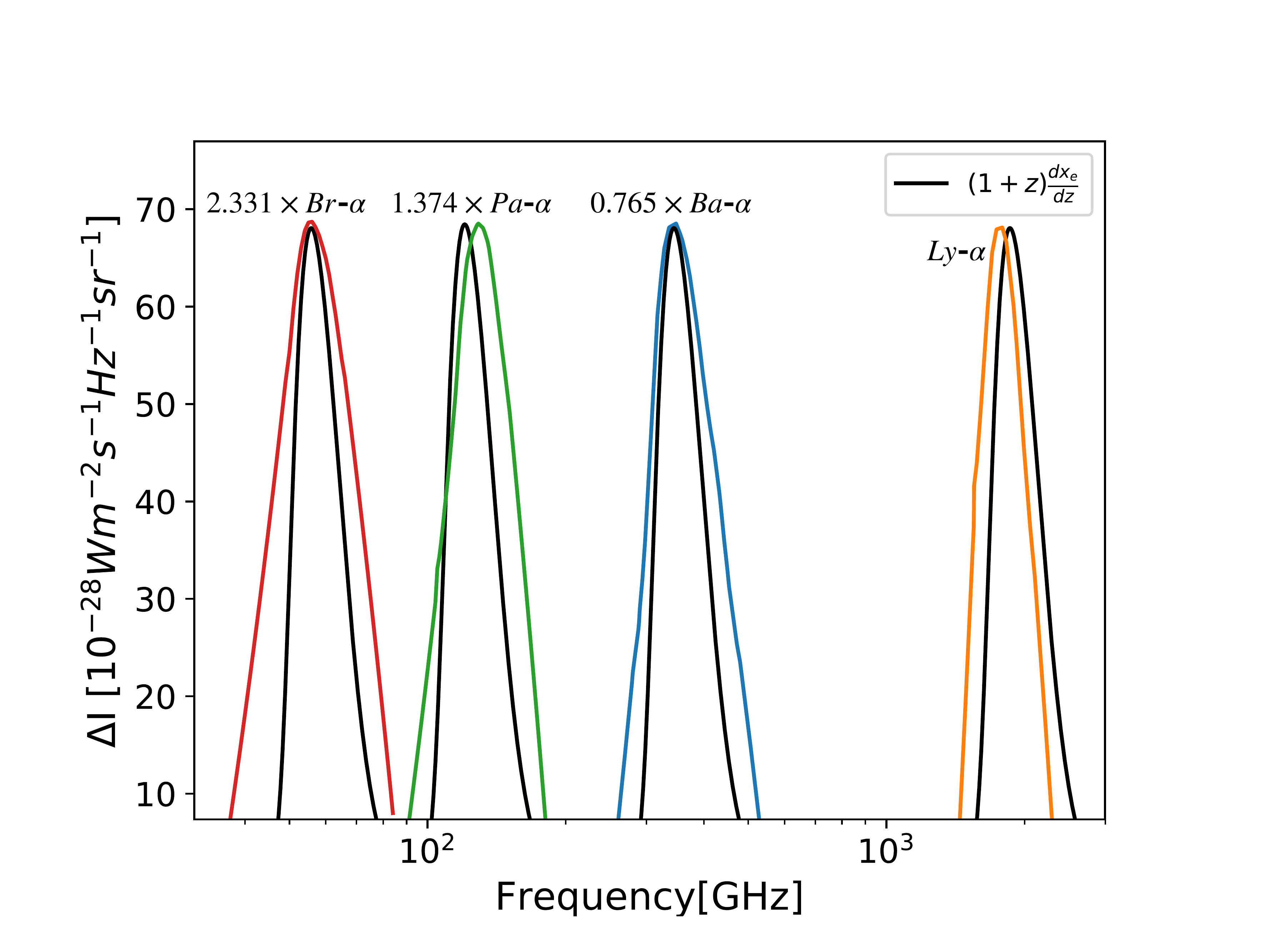}
 \caption{Comparison of our simple ansatz for the shape of the
   recombination lines (Eq. \ref{Eq:ansatz}) with the actual line shapes.}\label{fig:linecomp}
\end{figure}

The rate of change of electron fraction $\xe$ and recombination spectrum
are expected to be closely related to each other. We would expect in
general  that a
faster recombination at a particular redshift would produce more
recombination photons for every line at the corresponding redshifted
observed frequency. This leads us to postulate a simple ansatz for the
shape of the recombination lines. Let us assume that the net transition
rate per unit volume $\Delta R(z)$ at any line is just proportional to the recombination
rate, $\Delta R(z)\propto n_{\rm H}\frac{d\xe}{dt}$, where $n_{\rm H}$ is
the total hydrogen number density.  The intensity in a
recombination line is given by \cite{Wong:2005yr,rcs2006}
\barr
I_\nu=\frac{ch}{4\pi}\frac{\Delta R(z)}{H(1+z)^3}\\
\earr
Putting $\Delta R(z)\propto n_{\rm H}\frac{d\xe}{dt}$, the line shape is given by
\barr
I_\nu=A (1+z)\frac{d\xe}{dz},\label{Eq:ansatz}
\earr
where $A$ is the normalization which is different for every line. We
compare the intensity calculated with this ansatz, after fitting the
amplitude $A$ so that the lines have the same normalization, in Fig \ref{fig:linecomp}.
Note that  we need to relate the observed frequency with  the redshift as 
\barr
(1+z)=\frac{\nu_{em}}{\nu_{obs}},
\earr
where, $\nu_{em}$ and $\nu_{obs}$ are the emitted and observed frequencies of the line  respectively.
We see from Fig. \ref{fig:linecomp} that the above very simple ansatz gets
the peak of line position right for the Balmer and Brackett lines, and
quite close for the Lyman and Paschen lines. The line width is close for
the  Lyman-$\alpha$ line but underestimated by our ansatz for the other
lines. 
However in general the line shapes are
similar enough and this gives support to  the idea that the recombination spectrum
carries detailed information about the whole history of recombination
compared to the integrated effect of the visibility function that we are
sensitive to with the CMB anisotropies. This is particularly important if
we want to constrain new physics which modifies the recombination history
but without significant effect on the visibility function. We leave
exploration of such scenarios to future work.

\section{Conclusions}
Cosmological recombination spectrum promises to  provide an exciting  new
window to look at the early universe, one that is not limited by cosmic
variance and can therefore result in unprecedented precision for the
measurement of standard cosmological parameters. This
will be an independent measurement of the cosmological
parameters. Improvement of precision in independent probes is of tremendous
importance, as it may uncover new anomalies which may point to the new
physics beyond the standard model.

The quantification of the information content of the cosmological
recombination spectrum and how to extract it from data in the presence of
foreground, are both very important questions for building the science case
and designing future missions \citep{Chluba_et_al_2019_vision_2050}. We have tried to answer the first question in this paper and leave
exploration of the second question to future work.

 \section*{Acknowledgments}
Debajyoti Sarkar  would like to thank  Tarun Souradeep for encouragement
and comments on the work. This work was supported by Science and
Engineering Research Board, Department of Science and Technology, Govt. of
India grant SERB/ECR/2015/000078 and Max Planck partner group between Tata
Institute of Fundamental Research, Mumbai and Max Planck Institute for
Astrophysics, Garching funded by Max-Planck-Gesellschaft.

\bibliographystyle{JHEP}
\bibliography{recom}
\end{document}